\documentclass[letterpaper, 10 pt, conference]{ieeeconf}  % Comment this line out if you need a4paper

\IEEEoverridecommandlockouts                              % This command is only needed if 
\usepackage{graphics} % for pdf, bitmapped graphics files
\usepackage{epsfig} % for postscript graphics files
\usepackage{mathptmx} % assumes new font selection scheme installed
\usepackage{times} % assumes new font selection scheme installed
\usepackage{amsmath} % assumes amsmath package installed
\usepackage{amssymb}  % assumes amsmath package installed
\usepackage{cite}

\title{\LARGE \bf
Multi-feature Multi-Scale CNN-Derived COVID-19 Classification from Lung Ultrasound Data
}

\author{Hui Che, Jared Radbel, Jag Sunderram, John L. Nosher, Vishal M. Patel, and Ilker Hacihaliloglu % <-this % stops a space
\thanks{Hui Che is with the Department of Biomedical Engineering, Rutgers University, NJ, USA}
\thanks{Jared Radbel and Jag Sunderam are with the Department of Medicine, Rutgers Robert Wood Johnson Medical School, NJ, USA}
\thanks{John L. Nosher is with the Department of Radiology, Rutgers Robert Wood Johnson Medical School, NJ, USA}
\thanks{Vishal M. Patel is with the Department of Electrical and Computer Engineering, Johns Hopkins University, MD, USA}
\thanks{Ilker Hacihaliloglu is with the Department of Biomedical Engineering, Rutgers University and the Department of Radiology, Rutgers Robert Wood Johnson Medical School, NJ, USA (Corresponding author; E-mail: \tt\small ilker.hac@soe.rutgers.edu )}}

\usepackage{multirow}

\begin{document}

\maketitle
\thispagestyle{empty}
\pagestyle{empty}

%%%%%%%%%%%%%%%%%%%%%%%%%%%%%%%%%%%%%%%%%%%%%%%%%%%%%%%%%%%%%%%%%%%%
\begin{abstract}

The global pandemic of the novel coronavirus disease 2019 (COVID-19) has put  tremendous  pressure  on  the  medical  system.  Imaging  plays  a  complementary  role   in the management of patients with COVID-19. Computed tomography (CT) and chest X-ray (CXR) are the two dominant screening tools. However, difficulty in eliminating the risk of disease transmission, radiation exposure and not being cost- effective are some of the challenges for CT and CXR imaging. This fact induces the implementation of lung ultrasound (LUS) for evaluating COVID-19 due to its practical advantages of noninvasiveness, repeatability, and sensitive bedside property. In this paper, we utilize a deep learning model to perform the classification of COVID-19 from LUS data, which could produce objective diagnostic information for clinicians. Specifically, all LUS images are processed to obtain their corresponding local phase filtered images and radial symmetry transformed images before fed into the multi-scale residual convolutional neural network (CNN). Secondly, image combination as the input of the network is used to explore rich and reliable features. Feature fusion strategy at different  levels  is  adopted  to  investigate  the relationship between the depth of feature aggregation and the classification accuracy. Our proposed method is evaluated on the point-of-care US (POCUS) dataset together with the Italian COVID-19 Lung US database (ICLUS-DB) and shows promising performance for COVID-19 prediction.
%\newline

%\indent \textit{Clinical relevance}— This is a brief additional statement on why a this might be of interest to %practicing clinicians. Example: This establishes the anesthetic efficacy of 10\% intraosseous injections with %epinephrine to positively influence cardiovascular function.
\end{abstract}

%%%%%%%%%%%%%%%%%%%%%%%%%%%%%%%%%%%%%%%%%%%%%%%%%%%%%%%%%%%%%%%%%%%%%%%%%%%%%%%%
\section{INTRODUCTION}

The COVID-19 pandemic has increased the burden of excess morbidity and mortality worldwide. The high trans- missibility and long incubation time of the SARS CoV-2 virus increase the difficulty in containing viral spread. A rapid diagnosis and severity classification in the early stage of the disease can significantly reduce the risk of further infections and help mitigate the excess morbidity and mor- tality of critically ill patients. At present, the main detection methods for COVID-19 infection are the genetic test (reverse transcription polymerase chain reaction (RT-PCR)) \cite{c1} and serology test. RRadiological assessment, based on CT and CXR, has been incorporated to improve the management of COVID-19 disease. However, difficulty in eliminating the risk of disease transmission, radiation exposure, not being cost-effective are some of the challenges for CT and CXR imaging \cite{c2}. CT scan can also be not performed bedside limiting its use in the intensive care unit (ICU) settings.

Lung ultrasound (LUS) is non-invasive, rapid, repeatable, and provides bedside imaging providing a safer alternative  to CXR and CT. As such, LUS use for rapid assessment of the severity of COVID-19 pneumonia has been reported \cite{c1, c2, c3}. However, early lesions or less obvious tissue changes are difficult to distinguish by the human eyes. Furthermore, differences in medical pathology around various regions and the varied LUS experience of clinicians can result in misdiagnosis. Thus, developing standardized systems to report and interpret disease findings is a challenge with LUS  \cite{c2}.

Artificial intelligence (AI)-based solutions in medical imaging have demonstrated the potential to establish objective and unified interpretation standards. In \cite{c1} a new convolutional neural network (CNN) architecture, termed POCOVID-Net, was proposed. A VGG-16 architecture was used as the backbone and was fine-tuned during network training \cite{c1}. The reported average 3-class classification accuracy was 89\% \cite{c1}. In \cite{c4}, a multi-task CNN architecture was proposed. The network achieved an $F1_{score}$ of 61\%, a precision of 70\%, and a recall of 60\% for risk prediction.

Although promising early results, CNN-based methods for processing B-mode US data are affected by the image acquisition settings and quality of the collected data \cite{c5}. Finally, the limited availability of COVID-19 LUS data is also another bottleneck. 

To address the above problems, we propose using a multi-feature multi-scale CNN-based approach to achieve a more accurate COVID-19 classification. Given that incorporating local-phase image tissue features can improve the accuracy of CNNs\cite{c5} for processing B-mode US data, local phase US image-based COVID-19 signatures are extracted for diverse and robust representations. Then we adopted the feature-fusion strategy to realize the effect of feature complement. To enlarge the network perception dimensions for more discriminative features of the input images, extra convolutional layers with different-size kernels are used in our CNN architecture. Our proposed approach is evaluated on 1752 scans obtained from 76 subjects.

\section{METHODS}

Our method mainly consists of two parts, local phase features extraction and binary classification based on multi-feature multi-scale CNNs. In this work, the use of local phase information aims to enhance the appearance of lung tissues and recover the pertinent tissue structure from LUS data. The extraction of local phase image features also increase the dataset size for training. The model applied for the classification task is based on the multi-scale two-dimensional (2D) residual neural network (ResNet) architecture similar to the one reported in \cite{c6}. Three different fusion architectures are investigated during this work.

\subsection{Local Phase Image Features}

Image phase information is a key component in the interpretation of a scene and has been used in various applications for processing US data \cite{c5,c7}. In this part, we first obtain the local phase energy feature image, denoted as $LPE(x, y)$, which is extracted using:
\begin{eqnarray}
LPE(x, y) = \sum_{sc}\left| US_{M1} \right| - \sqrt{US_{M2}^2 + US_{M3}^2}
\end{eqnarray}
In the above formula, $sc$ represents the number of filter scales set to 2 throughout the experimental evaluation, and $US_{M}$ is the group of monogenic signal images computed using the vector-valued odd filter (Riesz filter) \cite{c7} on band-pass filtered LUS image, denoted as $US_{B}(x,y)$, as follows:
\begin{eqnarray}
\begin{aligned}
US_{M}(x, y)  =[&US_{M1}(x, y), US_{M2}(x, y), US_{M3}(x, y)]\\
=[&US_{B}(x, y), US_{B}(x, y) \otimes h_{1}(x, y),\\
&US_{B}(x, y) \otimes h_{2}(x, y)]
\end{aligned}
\end{eqnarray}
where $\otimes$ represents the convolution operation and $h_{1}(x,y)$, $h_{2}(x,y)$ are components in Riesz filter. For bandpass filtering $\alpha$-scale space derivative quadrature filters (ASSD) \cite{c7} are used to output $US_{B}(x,y)$.

Then US signal transmission map is modelled with scattering and attenuation information to get enhanced image $US_{E}(x, y)$, with maximized visibility of high intensity $LPE(x, y)$ features inside a local region. A linear interpolation model is selected to combine the two interactions:
\begin{eqnarray}
LPE(x,y) = US_{A}(x,y)US_{E}(x,y) + (1 - US_{A}(x,y))\beta
\end{eqnarray}
Here, $LPE(x, y)$ is the local phase energy image, $US_{A}(x, y)$ is the signal transmission map and $US_{E}(x, y)$ is the enhanced image. $\beta$ is a constant value representative of the tissue echogenicity in the local region. Our aim is $US_{E}(x, y)$ and we hope to get two different enhancement results with different $\beta$ value settings (60\% and 90\% of the maximum intensity value of $LPE(x,y)$). Once the signal transmission map $US_{A}(x, y)$ is obtained using the well-established Beer Lambert Law, $US_{E}(x, y)$ can be calculated according to Equation (4):
\begin{eqnarray}
US_{E}(x, y) = \frac{LPE(x, y) - \beta}{[max(US_{A}(x, y), \epsilon)]^\delta} + \beta 
\end{eqnarray}
$\delta$ is related to tissue attenuation coefficient, $\eta$, and $\epsilon$ is a small constant to avoid division by zero. Throughout the experimental evaluation $\eta=0.85$, $\epsilon=0.0001$. ASSD filter parameters were kept same as reported in \cite{c7}.

In Fig. 1, the two enhanced local phase images $US_{E}(x, y)$ denoted as $US_{E1}(x, y)$ and $US_{E2}(x, y)$ are shown. 
These two enhanced images are used as the input for radial symmetry tissue extraction.
Fast radial symmetry transform algorithm is applied on the local phase images, aiming to detect points of interest \cite{c7}. Fig. 1 shows radial symmetry images $S_{1}(x, y)$ and $S_{2}(x, y)$ corresponding to the local phase enhanced images, $US_{E1}(x, y)$ and $US_{E2}(x, y)$, it can be seen that the transformation highlights the points of interest that are characterized by radial symmetry as well as high contrast.

\begin{figure}[!t]
  \centering
  \includegraphics[width=8.5cm]{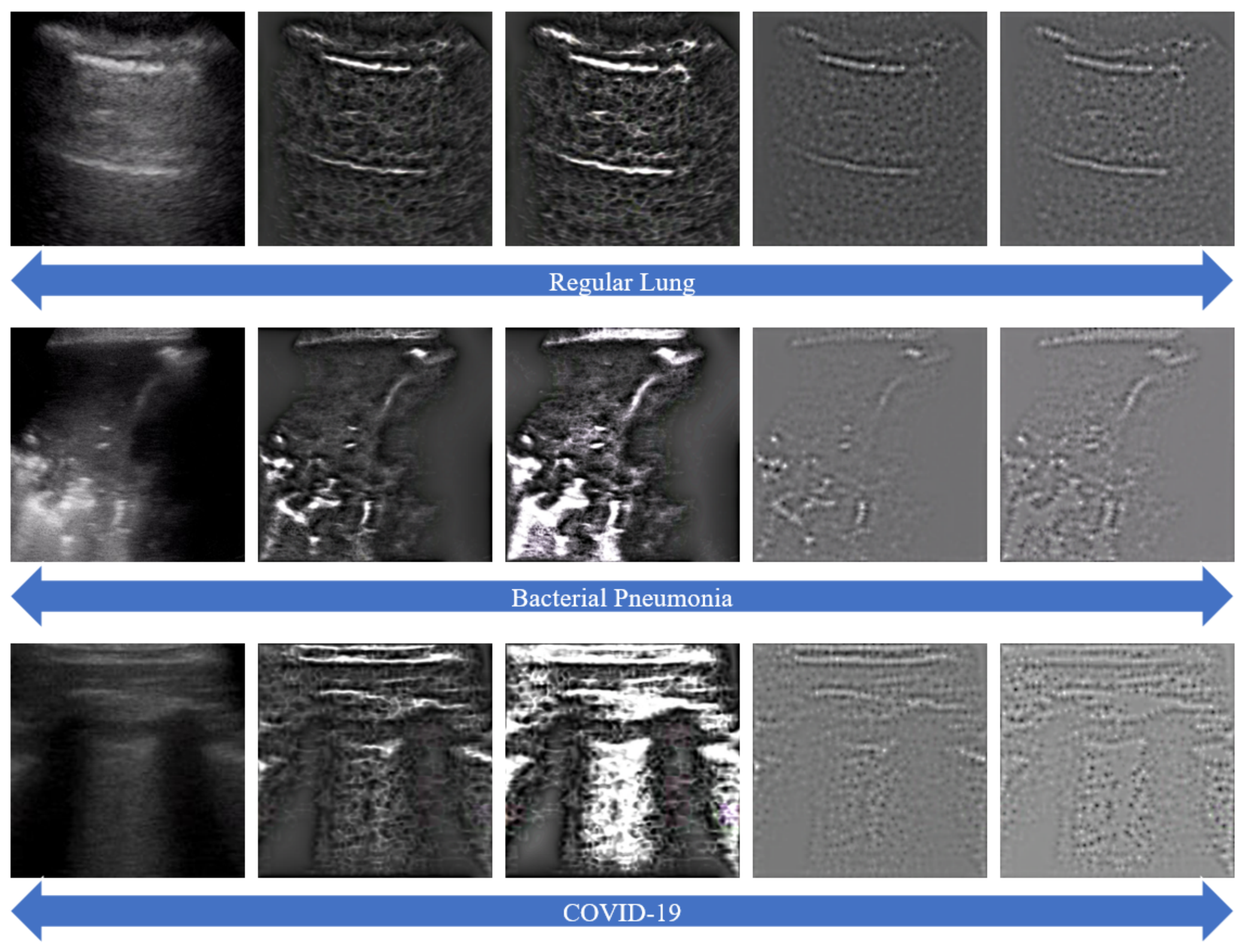}
%  \vspace{2.0cm}
%
  \caption{Qualitative results of local phase and radial symmetry-based image enhancement and feature extraction methods. Top row: A regular lung. Middle row: A bacterial pneumonia infected lung. Bottom row:A COVID-19 infected lung. All rows from left to right: LUS image $US(x, y)$, local phase enhanced images $US_{E1}(x, y)$ and $US_{E2}(x, y)$, radial symmetry transformed images $S_{1}(x, y)$ and $S_{2}(x, y)$.}
  \label{figurelabel}
\end{figure}

\subsection{Network Architecture}

The multi-scale 2D ResNet is a light-weighted classification network even though it simultaneously captures features from multiple receptive fields. 
This network is composed of three functional parts: 1) one convolutional layer for primary feature map extraction, 2) multiple residual blocks with the multi-scale convolutional layer, 3) a fully connected layer with softmax activation function to act as a classifier. 
All input images should be resized to $512\times512$ before fed into the network.
We investigate three different fusion architectures with the $US_(x, y)$, $US_{E1}(x, y)$, $US_{E2}(x, y)$, $S_{1}(x, y)$, and $S_{2}(x, y)$ images as an input. 

\begin{figure*}[!t]
\centering
\centerline{\includegraphics[width=17cm]{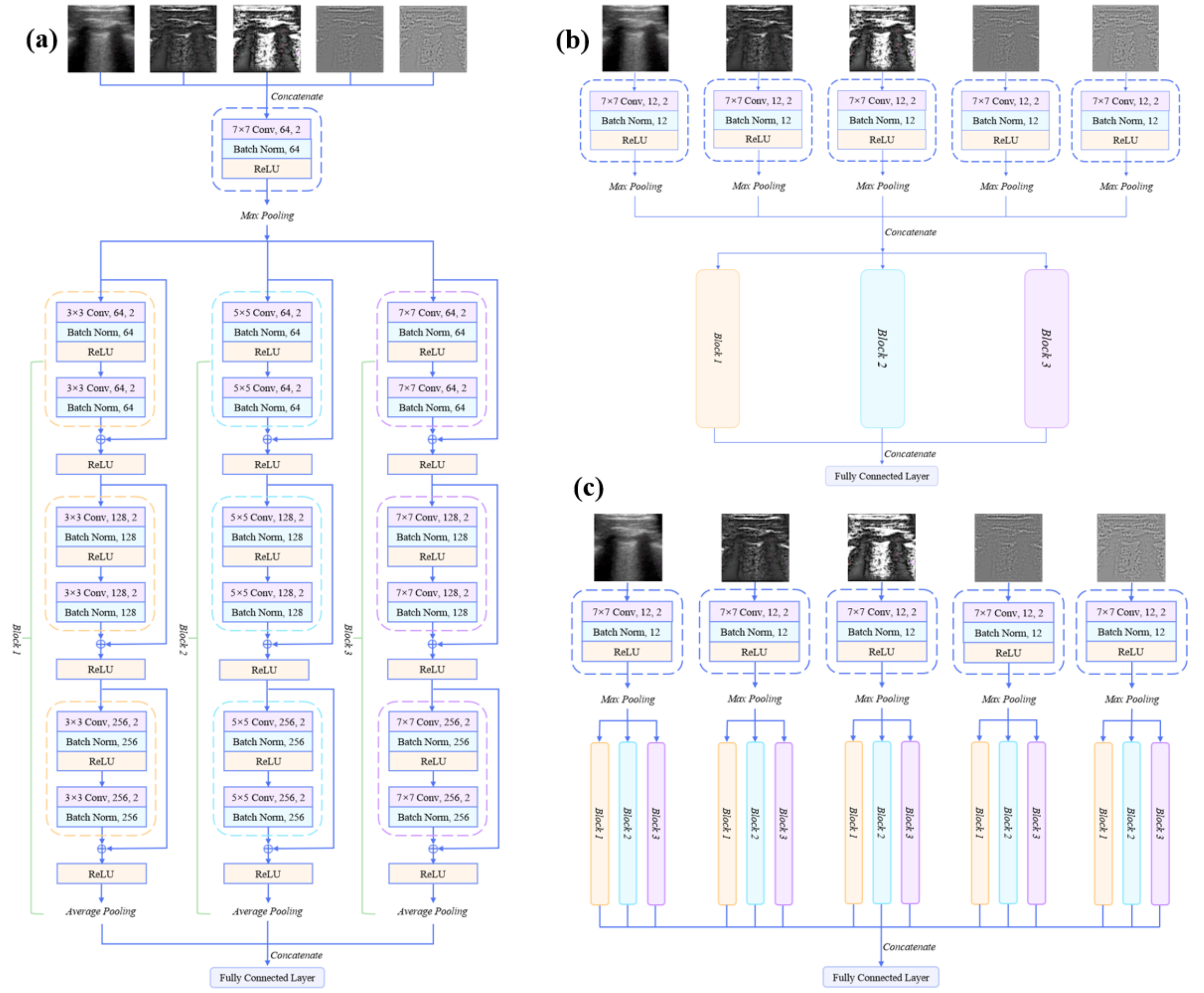}}
\caption{The various CNN architectures. Each convolutional layer has three parameters: kernel size, depth and stride. (a) The architecture of the early-fusion CNN. $US(x,y)$, $US_{E1}(x,y)$, $US_{E2}(x,y)$, $S_{1}(x,y)$ and $S_{2}(x,y)$ images are fused at the pixel level to input the network. (b) The architecture of the mid-fusion CNN. Five input images are separately processed by the initial convolutional block to output the corresponding primary feature maps. A concatenation of these primary feature maps is processed through the network. (c) The architecture of the late-fusion CNN. Five input images are separately processed through the whole network till the average pooling layer. All final feature maps are fused to input the fully-connected layer.}
%\label{fig:res}
\end{figure*}

Fig. 2 illustrates various network architectures. Since the scale of the objects in the image is varies, we adopt multi-scale receptive fields to focus on different scale information.  
As shown in Fig. 2, three ResCNN blocks are basic components in all designs to extract multi-scale features. They have different receptive kernels, the sizes set to $3 \times 3$, $5 \times 5$ and $7 \times 7$. In every ResCNN block, there are three sub-blocks and each sub-block contains two convolutional layers. The skip connection is added in each sub-block to avoid the degradation problem \cite{c5}. An average pooling layer follows after the convolution operation to output the final feature map. At the end of the network, a fully connected layer with activation function is used to act as a classifier, with the input of the concatenation of the final feature maps.

Feature-fusion function is utilized in different levels of the CNN model to construct early-, mid-, and late-fusion structures \cite{c4}. To achieve early-fusion, all the images are concatenated at the pixel level to form input with more channels. In the mid-fusion model, multiple input images are input to the network separately, processed by the initial convolutional layer to obtain corresponding primary feature maps. Concatenation is performed to finish feature aggregation for the deeper extraction. Late fusion is operated before the fully connected layer processing to fuse final feature maps from each input image. 

\subsection{Data} 
The dataset used in this work was obtained from \cite{c1} and \cite{c3} and consisted of 1276 COVID-19 LUS scans from 51 subjects, 254 bacterial pneumonia LUS scans from 13 subjects, and 222 LUS scans from 12 healthy subjects.
All selected images from two released datasets are in the form of convex probe image. Bacterial pneumonia and healthy LUS scans are joined as a non-COVID-19 class. Our task is to perform binary classification.
Before image enhancement, all data is cropped into $334 \times 334$ squares for a purpose of removing the non-relevant information.

\begin{table*}
	\caption{Classification performance summary. Best result is shown in \textbf{bold}}
	\label{tab:my_label}
	\centering
	\begin{tabular}{|l|l|l|l|l|l|} 
		\hline
		\textbf{Method}                                                      &                                                                               & Accuracy                                                                     & \begin{tabular}[c]{@{}l@{}}Precision\\Covid-19/Non\end{tabular}                                       & \begin{tabular}[c]{@{}l@{}}\\Recall\\Covid-19/Non\\\end{tabular}                                            & \begin{tabular}[c]{@{}l@{}}\\F1 Score\\Covid-19/Non\\\end{tabular}                                     \\ 
		\hline
		$US(x,y)$(single feature CNN)                                         &                                                                               & 89.94\%                                                                      & 92.48\%/82.49\%                                                                                       & 93.98\%/78.70\%                                                                                             & 93.21\%/80.46\%                                                                                        \\ 
		\hline
		$US_{E1}(x,y)+US_{E2}(x,y)$                                          & \begin{tabular}[c]{@{}l@{}}Early Fusion\\Mid-Fusion\\Late-Fusion\end{tabular} & \begin{tabular}[c]{@{}l@{}}91.93\%\\90.91\%\\88.96\%\end{tabular}            & \begin{tabular}[c]{@{}l@{}}\textbf{95.18\%}/84.65\%\\94.94\%/81.69\%\\92.12\%/81.29\%\end{tabular}    & \begin{tabular}[c]{@{}l@{}}93.76\%/\textbf{87.40\%}\\\textbf{}92.73\%/86.80\%\\92.99\%/78.72\%\end{tabular} & \begin{tabular}[c]{@{}l@{}}94.39\%/85.52\%\\93.72\%/83.56\%\\92.44\%/79.39\%\end{tabular}              \\ 
		\hline
		$S_{1}(x,y)+S_{2}(x,y)$                                              & \begin{tabular}[c]{@{}l@{}}Early Fusion\\Mid-Fusion\\Late-Fusion\end{tabular} & \begin{tabular}[c]{@{}l@{}}90.68\%~ ~\\86.53\%~\\87.52\%~ ~ ~\end{tabular}   & \begin{tabular}[c]{@{}l@{}}93.09\%/84.35\%\\87.35\%/84.92\%\\89.70\%/83.38\%\end{tabular}             & \begin{tabular}[c]{@{}l@{}}94.33\%/80.85\%\\95.85\%/61.54\%\\93.62\%/70.46\%\end{tabular}                   & \begin{tabular}[c]{@{}l@{}}93.65\%/82.17\%\\91.27\%/69.50\%\\91.48\%/75.56\%\end{tabular}              \\ 
		\hline
		$US(x,y)+ US_{E1}(x,y)+US_{E2}(x,y)$                                 & \begin{tabular}[c]{@{}l@{}}Early Fusion\\Mid-Fusion\\Late-Fusion\end{tabular} & \begin{tabular}[c]{@{}l@{}}88.60\%\\92.80\%~\\\textbf{95.11\%}~\end{tabular} & \begin{tabular}[c]{@{}l@{}}93.01\%/78.21\%\\93.28\%/91.74\%\\94.93\%/\textbf{95.87\%~} ~\end{tabular} & \begin{tabular}[c]{@{}l@{}}91.42\%/81.70\%\\97.18\%/81.54\%\\\textbf{98.59\%}/86.05\%\end{tabular}          & \begin{tabular}[c]{@{}l@{}}92.09\%/79.27\%\\95.14\%/86.03\%\\\textbf{96.70\%/90.48\%~} ~\end{tabular}  \\ 
		\hline
		$US(x,y)+ US\_\{E1\}(x,y)+US\_\{E2\}(x,y)+S\_\{1\}(x,y)+S\_\{2\}(x,y)$ & \begin{tabular}[c]{@{}l@{}}Early Fusion\\Mid-Fusion\\Late-Fusion\end{tabular} & \begin{tabular}[c]{@{}l@{}}90.57\%~\\88.79\%~\\89.33\%~\end{tabular}         & \begin{tabular}[c]{@{}l@{}}93.51\%/84.21\%\\92.37\%/79.50\%~\\92.26\%/82.15\%~\end{tabular}           & \begin{tabular}[c]{@{}l@{}}93.89\%/82.74\%\\92.55\%/79.20\%~\\93.25\%/79.05\%~\end{tabular}                 & \begin{tabular}[c]{@{}l@{}}93.54\%/82.50\%\\92.37\%/78.80\%~\\92.69\%/80.16\%~\end{tabular}            \\
		\hline
	\end{tabular}
\end{table*}

\subsection{Experiment Implementation}
We perform 5-fold cross-validation to evaluate the performance of our proposed method. During evaluation same patient data was not included in the training and testing data. The reported final results show the mean of the 5-fold cross validation. All datasets maintain the same data distribution, including the $US(x, y)$ dataset and $US_{E1}(x,y)$, $US_{E2}(x,y)$, $S_{1}(x,y)$, $S_{2}(x,y)$ datasets. 

All CNN models are trained by using the cross-entropy loss function and ADAM optimizer with a learning rate of $1e-5$. Classification performance is measured by four metrics: accuracy, precision, recall and $F1_{score}$. 
To evaluate the effectiveness of image processing methods and feature-fusion strategies, we compare the results of just using $US(x, y)$ image as the input and the combination between two groups of processed images ($US_{E1}(x, y)$ and $US_{E2}(x, y)$, $S_{1}(x, y)$ and $S_{2}(x, y)$). Furthermore, we investigate the accuracy of the model by using 5 kinds of images as input.

Experiments are implemented in the PyTorch framework with an Intel Core GPU at 3.70 GHz and a NVIDIA GeForce GTX 1080Ti GPU. 

\section{RESULTS}

%\begin{itemize}

%\item Use either SI (MKS) or CGS as primary units. (SI units are encouraged.) English units may be %used as secondary units (in parentheses). An exception would be the use of English units as %identifiers in trade, such as Ò3.5-inch disk driveÓ.
%\item Avoid combining SI and CGS units, such as current in amperes and magnetic field in oersteds. %This often leads to confusion because equations do not balance dimensionally. If you must use mixed %units, clearly state the units for each quantity that you use in an equation.
%\item Do not mix complete spellings and abbreviations of units: ÒWb/m2Ó or Òwebers per square meterÓ, %not Òwebers/m2Ó.  Spell out units when they appear in text: Ò. . . a few henriesÓ, not Ò. . . a few %HÓ.
%\item Use a zero before decimal points: Ò0.25Ó, not Ò.25Ó. Use Òcm3Ó, not ÒccÓ. (bullet list)

%\end{itemize}

Quantitative results of our proposed method are presented in Table 1. Our proposed multi-scale network achieves an average classification accuracy of 89.94\% when using only LUS data ($US(x,y)$). The average accuracy increases to 91.93\% when using enhanced images $US_{E}(x,y)$, and 90.68\% when using radial symmetry images $S(x,y)$ (Table 1). The best performance was obtained when combining the LUS $US(x,y)$ images with the enhanced images $US_{E}(x,y)$, where an average accuracy of 95.11\% was obtained. The comparison among results of the first three sets of experiments demonstrates that local phase feature is beneficial to enhance tissue characteristics for network learning, especially, feature fusion performed in the early stage. As seen in Table 1 the results present that late-fusion design obtains the highest accuracy (95.11\%), $F1_{score}$ (96.70\%) significantly outperforming the other fusion operations in these two metrics. When all the image features were combined early fusion architecture obtained the best results compared to other fusion networks investigated. We further observe that by using local phase image features the performance of the network for classifying non-COVID-19 data is also improved ($F1_{score}$ of 90.48\% vs 80.46\%).

\section{CONCLUSION}

We proposed to apply a novel CNN-based method to achieve accurate COVID-19 prediction from LUS. Quantitative and qualitative results confirm that the use of local phase information and multi-feature multi-scale CNN contributes to improved COVID-19 classification performance in LUS data. Fusing LUS features and local phase features at a late stage gives the highest accuracy reaching 95.11\%, at the same time, other metrics prove a balanced classification capability of the model. In most cases, early-fusion strategy shows a better classification performance. Our future work will include the evaluation of the proposed method on a larger scale dataset. We also would like to extend our network for multi-class classification for differentiating regular pneumonia from COVID-19. Finally, optimization of the local phase image filter parameters based on CNN performance will be another future work.

%\section{CONCLUSIONS}

%A conclusion section is not required. Although a conclusion may review the main points of the paper, %do not replicate the abstract as the conclusion. A conclusion might elaborate on the importance of the %work or suggest applications and extensions. 


%%%%%%%%%%%%%%%%%%%%%%%%%%%%%%%%%%%%%%%%%%%%%%%%%%%%%%%%%%%%%%%%%%%%%%%%%%%%%%%%

%%%%%%%%%%%%%%%%%%%%%%%%%%%%%%%%%%%%%%%%%%%%%%%%%%%%%%%%%%%%%%%%%%%%%%%%%%%%%%%%

%%%%%%%%%%%%%%%%%%%%%%%%%%%%%%%%%%%%%%%%%%%%%%%%%%%%%%%%%%%%%%%%%%%%%%%%%%%%%%%%

%%%%%%%%%%%%%%%%%%%%%%%%%%%%%%%%%%%%%%%%%%%%%%%%%%%%%%%%%%%%%%%%%%%%%%%%%%%%%%%%


\begin{thebibliography}{99}

\bibitem{c1} Born J, Brändle G, Cossio M, et al. POCOVID-Net: automatic detection of COVID-19 from a new lung ultrasound imaging dataset (POCUS)[J]. arXiv preprint arXiv:2004.12084, 2020.

\bibitem{c2} Schultz M J, Sivakorn C, Dondorp A M. Challenges and opportunities for lung ultrasound in novel coronavirus disease (COVID-19)[J]. The American Journal of Tropical Medicine and Hygiene, 2020, 102(6): 1162.

\bibitem{c3} Soldati G, Smargiassi A, Inchingolo R, et al. Proposal for International Standardization of the Use of Lung Ultrasound for Patients With COVID‐19: A Simple, Quantitative, Reproducible Method[J]. Journal of Ultrasound in Medicine, 2020.

\bibitem{c4} Roy S, Menapace W, Oei S, et al. Deep learning for classification and localization of COVID-19 markers in point-of-care lung ultrasound[J]. IEEE Transactions on Medical Imaging, 2020.

\bibitem{c5} Alsinan A Z, Patel V M, Hacihaliloglu I. Automatic segmentation of bone surfaces from ultrasound using a filter-layer-guided CNN[J]. International journal of computer assisted radiology and surgery, 2019, 14(5): 775-783.

\bibitem{c6} Liu R, Wang F, Yang B, et al. Multiscale Kernel Based Residual Convolutional Neural Network for Motor Fault Diagnosis Under Nonstationary Conditions[J]. IEEE Transactions on Industrial Informatics, 2019, 16(6): 3797-3806.

\bibitem{c7} Hacihaliloglu I. Localization of bone surfaces from ultrasound data using local phase information and signal transmission maps[C]. International Workshop and Challenge on Computational Methods and Clinical Applications in Musculoskeletal Imaging. Springer, Cham, 2017: 1-11.

\bibitem{c8} Loy G, Zelinsky A. Fast radial symmetry for detecting points of interest[J]. IEEE Transactions on pattern analysis and machine intelligence, 2003, 25(8): 959-973.

\end{thebibliography}
\end{document}